\documentclass[a4paper,10pt,twoside]{cpc-hepnp}

\usepackage{multicol}
\usepackage{graphicx}
\usepackage{booktabs}
\usepackage{amssymb,bm,mathrsfs,bbm,amscd}
\usepackage[tbtags]{amsmath}
\usepackage{lastpage}
\begin{document}
%\begin{CJK*}{GBK}{song}

%\fancyhead[c]{\small Chinese Physics C~~~Vol. xx, No. x (201x) xxxxxx}
%\fancyfoot[C]{\small xxxxxx-\thepage}

%\footnotetext[0]{Received xx March 201x}

\title{Investigation of the near-threshold cluster resonance in $^{14}\rm{C}$}

\author{%
    Hong-Liang Zang $^{1}$%
\quad Yan-Lin Ye $^{1;1)}$\email{yeyl@pku.edu.cn}%
\quad Zhi-Huan Li $^{1}$%
\quad Jian-Song Wang $^{2}$\\%
\quad Jian-Ling Lou $^{1}$
\quad Qi-Te Li $^{1}$%
\quad Yu-Cheng Ge $^{1}$%
\quad Xiao-Fei Yang $^{1}$\\%
\quad Jing Li $^{3}$%
\quad Wei Jiang $^{4}$%
\quad Jun Feng $^{1}$%
\quad Qiang Liu $^{1}$%
\quad Biao Yang $^{1}$\\%
\quad Zhi-Qiang Chen $^{1}$%
\quad Yang Liu $^{1}$%
\quad Hong-Yi Wu $^{1}$%
\quad Chen-Yang Niu $^{1}$\\%
\quad Chen-Guang Li $^{1}$%
\quad Chun-Guang Wang $^{1}$%
\quad Xiang Wang $^{1}$%
\quad Wei Liu $^{1}$\\%
\quad Jian Gao $^{1,5}$%
\quad Han-Zhou Yu $^{1}$%
\quad Jun-Bin Ma $^{2}$%
\quad Peng Ma $^{2}$%
\quad Zhen Bai $^{2}$\\%
\quad Yan-Yun Yang $^{2}$%
\quad Shi-Lun Jin $^{2}$%
\quad Fei Lu $^{6}$%
}

\maketitle

\address{%
$^1$ School of Physics and State Key Laboratory of Nuclear Physics and Technology, Peking University, Beijing, 100871, China\\
$^2$ Institute of Modern Physics, Chinese Academy of Science, Lanzhou 730000, China\\
$^3$ Argonne National Laboratory, 9700 S Cass Ave B109, Lemont, Illinois 60439, USA\\
$^4$ Institute of High Energy Physics, CAS, Beijing, 100049, China\\
$^5$ RIKEN Nishina Center, 2-1 Hirosawa, Wako, Saitama 351-0198, Japan\\
$^6$ Shanghai Institute of Applied Physics, Chinese Academy of Science, Shanghai 200000, China\\
}

\begin{abstract}
 An experiment for $p(^{14}\rm{C}$,$^{14}\rm{C}^{*}\rightarrow^{10}\rm{Be}+\alpha)\mathit{p}$ inelastic excitation and decay was performed in inverse kinematics at a beam energy of 25.3 MeV/u. A series of $^{14}\rm{C}$ excited states, including a new one at 18.3(1) MeV, were observed which decay to various states of the final nucleus of $^{10}\rm{Be}$. A specially designed telescope-system, installed around the zero degree, played an essential role in detecting the resonant states near the $\alpha$-separation threshold. A state at 14.1(1) MeV is clearly identified, being consistent with the predicted band-head of the molecular rotational band characterized by the $\pi$-bond linear-chain-configuration. Further clarification of the properties of this exotic state is suggested by using appropriate reaction tools.
\end{abstract}



%\footnotetext[0]{\hspace*{-3mm}\raisebox{0.3ex}{$\scriptstyle\copyright$}2013
%Chinese Physical Society and the Institute of High Energy Physics
%of the Chinese Academy of Sciences and the Institute
%of Modern Physics of the Chinese Academy of Sciences and IOP Publishing Ltd}%

\begin{multicols}{2}

\section{Introduction}

One of the intriguing and active research area in nowadays nuclear physics is to understand the cluster degree of freedom, upon which are forming the new effective interactions and the new exotic structures£¬ being conceptually different from those emerged from the single-particle picture. According to the well-known Ikeda diagram \cite{Ikeda1968}, cluster structures are enhanced in states close to the corresponding cluster-separation threshold. When moving away from the $\beta$-stability-line towards the neutron-drip-line, the valence neutrons may help to stabilize the multi-center nuclear molecular system with much more varieties of clustering configurations \cite{Oertzen2006}. So far the theoretical studies have been largely advanced ~\cite{Horiuchi2012, Zhu2013}, but the experimental identification of the cluster structure is still limited to the beryllium isotopes (two-center system) and some individual states in heavier nuclei (\cite{Horiuchi2012, Yang2014, Jiang2017} and references therein). The experimental difficulties come basically from the requirement of determining several complementary signatures such as the large moments of inertia, the strong cluster-decay branching-ratio (BR), the characteristic transition strength and so on ~\cite{Catford2013, Yang2014}.

In recent years the focus of the cluster studies has been gradually moving into the carbon isotopes (three-center system) (\cite{Li2017} and references therein). The well-established Hoyle state at 7.65 MeV in $^{12}\rm{C}$ \cite{Hoyle1954} has been identified as an $\alpha$-particle Bose-Einstein condensate (BEC) state ~\cite{Tohsaki2001}. W. von Oertzen \emph{et al.} have proposed the molecular rotational bands with linear-chain or triangular cluster-configurations in neutron-rich $^{14}$C nucleus ~\cite{Oertzen2004}. The latest antisymmetrized-molecular-dynamics (AMD) calculations have quantitatively predicted three types of molecular rotational bands, associated with the triangle-, $\pi$-bond linear-chain- and  $\sigma$-bond linear-chain-configurations, in $^{14}$C ~\cite{Suhara2010, Baba2016, Baba2017}. The $0^+$ band-heads of the latter two configurations are predicted to locate at $\sim$14 MeV and $\sim$22 MeV ~\cite{Baba2016}, just beyond the $^{10}$Be (0 MeV) + $\alpha$ separation threshold (12.01 MeV) and  beyond the $^{10}$Be (6 MeV) + $\alpha$ threshold ($\sim$18 MeV), respectively. The experimental investigations of these exotic molecular structures have been conducted very recently ~\cite{Freer2014, Fritsch2016, Li2017, Yamaguchi2017}. In Ref.~\cite{Li2017}, a state at 22.5 MeV in $^{14}\rm{C}$ was reported, which characterizes a $\sigma$-bond linear-chain molecular structure. The measurement was performed by using the silicon-strip detectors. So far, this kind of detection system has generated consistent results for a number of experiments ~\cite{Soic2003, Oertzen2004, Oertzen2006}. On the other hand, the resonances reported in Refs.\cite{Freer2014, Fritsch2016, Yamaguchi2017} were  measured by the resonant-scattering method with the thick gas target, and were tentatively related to the $\pi$-bond molecular configuration. The spin assignments in these works are not consistent with each other, and also contradict the previous measurements which used solid targets together with the silicon-strip detectors. Nevertheless, they all suggested, via the projection \cite{Freer2014, Fritsch2016} or the tentative spin-parity assignment \cite{Yamaguchi2017}, a $0^+$ band-head within the $13 \sim 15$ MeV energy range. However, all states in this energy range, observed in the previous inclusive measurements, have already been assigned nonzero spins with high certainty, except a hint of resonance at about 14 MeV which has not been physically investigated so far \cite{Price2007}.

In this article, we present a measurement of the $p(^{14}\rm{C}$,$^{14}\rm{C}^{*}\rightarrow^{10}\rm{Be}+\alpha)\mathit{p}$ inelastic excitation and decay. A specially designed 0-degree telescope was employed to increase the detection efficiency for resonant states near the $\alpha$-decay threshold ~\cite{Yang2014}. A firm identification of a state at 14.1(1) MeV was realized and its spin-assignment is discussed.

\section{Experiment}

\begin{center}
\includegraphics[width=6cm]{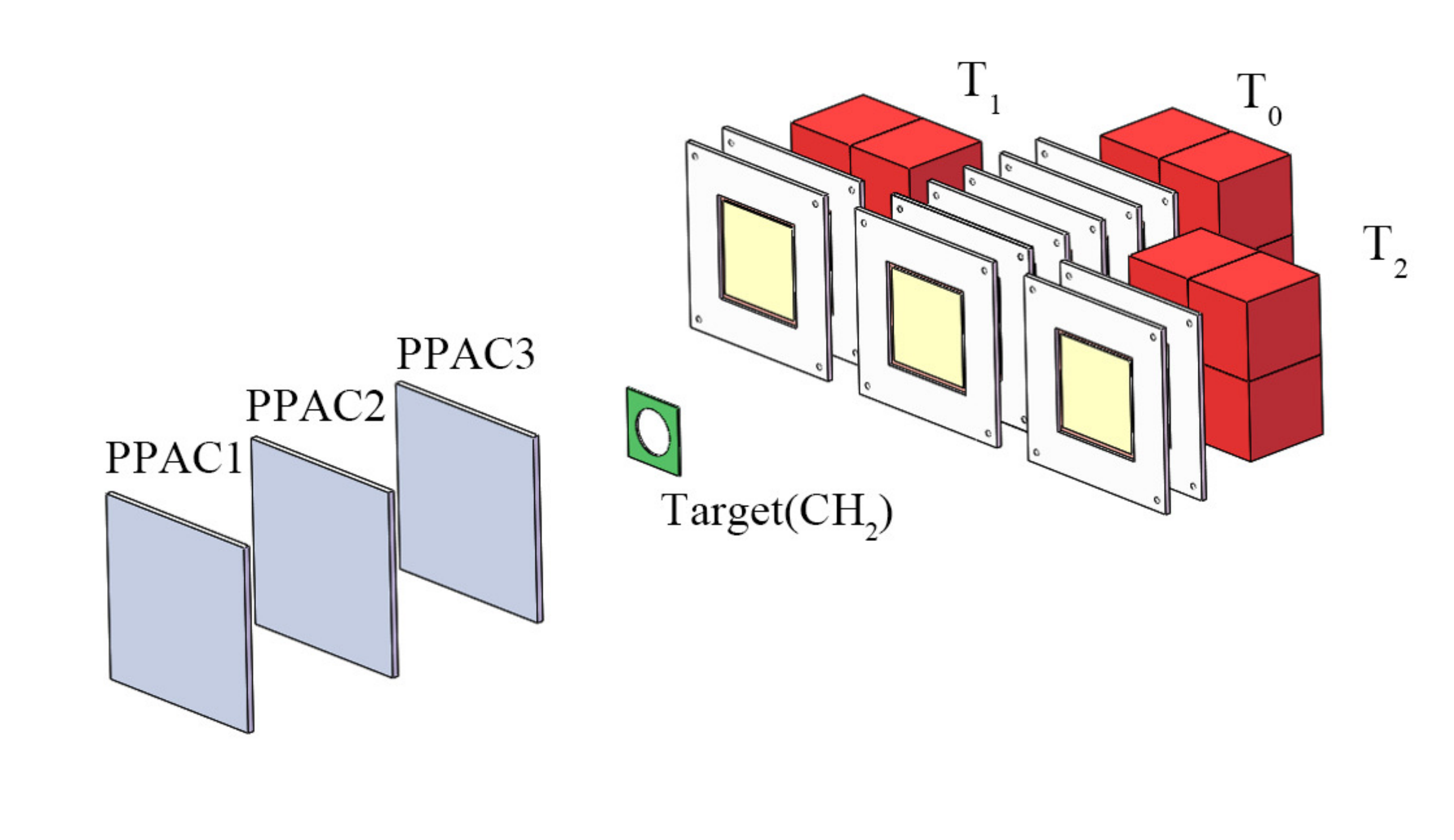}
\figcaption{\label{setup}A schematic view of the experimental setup.}
\end{center}

The experiment was carried out at the Radioactive Ion Beam Line at the Heavy Ion Research Facility in Lanzhou (HIRFL-RIBLL)\,\cite{Sun2003}. A 25.3 MeV/u $^{14}\rm{C}$ secondary beam with an intensity of $2\times10^4$ particle per second (pps) and a purity of 95\% was produced from an $^{18}\rm{O}$ primary beam at 70 MeV/u on a thick $^{9}\rm{Be}$ target. A schematic diagram of the detection system is displayed in Fig.~\ref{setup}. Three parallel plate avalanche chambers\,(PPAC) with position resolution of about 1 mm in both X and Y directions were employed to track the beam onto a 8.9 mg/$\rm{cm}^2$ $\rm{CH}_2$ target. The reaction products were detected by three telescopes, namely $\rm{T}_0$, $\rm{T}_1$ and $\rm{T}_2$. The forward moving charged fragments, such as the fragments from $^{14}\rm{C} \rightarrow ^{10}$Be + $\alpha$ breakup reaction, were recorded by the $\rm{T}_0$ telescope, which was centered at zero degree and located at a distance of 133 mm downstream of the target. The angular coverage of the $\rm{T}_0$ telescope was about $0^{\circ}-13.5^{\circ}$. The $\rm{T}_0$ telescope consisted of three double-sided silicon strip detectors (DSSD), three large-size silicon detectors (SSD) and a $2\times2$ CsI(Tl) scintillator array. The thickness of each DSSD or SSD is about 1000 $\mu$m or 1500 $\mu$m, respectively. The size of each CsI(Tl) unit is $41\times41\times40\ \rm{mm}^3$. The $\rm{T}_1$ and the $\rm{T}_2$ telescopes were aimed to detect the proton recoiled from the target and were centered at angles of $\pm~45^{\circ}$ relative to the beam direction. Each of them consisted of one DSSD with a thickness of 300 $\mu$m, one SSD with a thickness of 1500 $\mu$m and a $2\times2$ CsI(Tl) scintillator array. The active area of the DSSD or the SSD is about $64\times64\ \rm{mm}^2$. The front or back face of the DSSD is divided into 32 independent strips, with the direction of the strips on the front face being perpendicular to the ones on the back. Each CsI(Tl) scintillator unit is backed by a photodiode readout.

Using the standard method of energy-loss ($\Delta E$) versus remaining energy ($E_r$), the particle identification (PID) up to carbon isotopes was clearly distinguished, thanks to the outstanding energy resolution of the telescopes. The energy calibration of each silicon-detector was realized by using a combination of $\alpha$ sources. The $\Delta E$-$E_r$ back binding points of the PID spectra for various isotopes were also utilized for the energy calibration of the $\rm{T}_0$, which covered a broad energy range. The energy match for all silicon strips in one detector was achieved according to the uniform calibration method as described in Ref.\,\cite{Qiao2014}. The time information recorded from the strips were applied to exclude accidentally coincident events, which is of especial importance at a hitting-rate ($\rm{T}_0$) higher than $10^4$ Hz.

\section{Observed resonances and discussion}
In the present experiment, only events with double hits on the $\rm{T}_0$ or one hit on either the $\rm{T}_1$ or the $\rm{T}_2$ were recorded, according to the trigger-scheme of the data acquisition (DAQ) system. As the statistics of the triple coincident events were very low, energy and momentum conservations are used to provide a complete reconstruction of the reaction kinematics \cite{Tian2016,Li2017}. The momentum of the recoil proton is deduced from the difference between the momentum vector of the beam particle and that of the two detected fragments. Thus, the reaction $Q$-value can be calculated according to \cite{Tian2016,Li2017}:
\begin{eqnarray}
Q &=& E_{\rm{tot}}-E_{\rm{beam}} \nonumber \\
  &=& E_{^{10}\rm{Be}}+E_{\alpha}+E_{\rm{proton}}-E_{\rm{beam}}
\end{eqnarray}
Figure~\ref{Q} represents the experimental $Q$-value spectrum obtained from the coincident detection of $^{10}\rm{Be}$ and $\alpha$-particles in the T$_0$ telescope, assuming a proton recoil. The contamination from the carbon content in the CH$_2$ target has been eliminated by using the so-called EP-plot\,\cite{Costanzo-1990}. The spectrum is fitted with three Gaussian peaks. The highest $Q$-value peak ($Q_{\rm{ggg}}$ at about -12.0 MeV) corresponds to the reaction for all final particles on their ground states (g.s.). The peak at about -15.2 MeV is associated with $^{10}\rm{Be}$ in its first excited state (3.36 MeV, $2^+$). There is one more peak at about -17.9 MeV, corresponding to four adjacent states in $^{10}\rm{Be}$ at about 6 MeV excitation \cite{Soic2003, Li2017}. Owing to the much higher energy of the first excited state in $^{4}\rm{He}$ (20.21 MeV) and no observable excited state in proton, peaks at -15.2 and -17.9 MeV cannot correspond to the $^{4}\rm{He}$ or the proton excitation. The energy resolution (FWHM) of the Q-peaks is about 2.5 MeV, which is worse than the intrinsic energy resolution of the telescopes. This relatively poor resolution is primarily attributed to the energy spread of the secondary beam and the uncertainty in determining the reaction point on the target.
\begin{center}
\includegraphics[width=6cm]{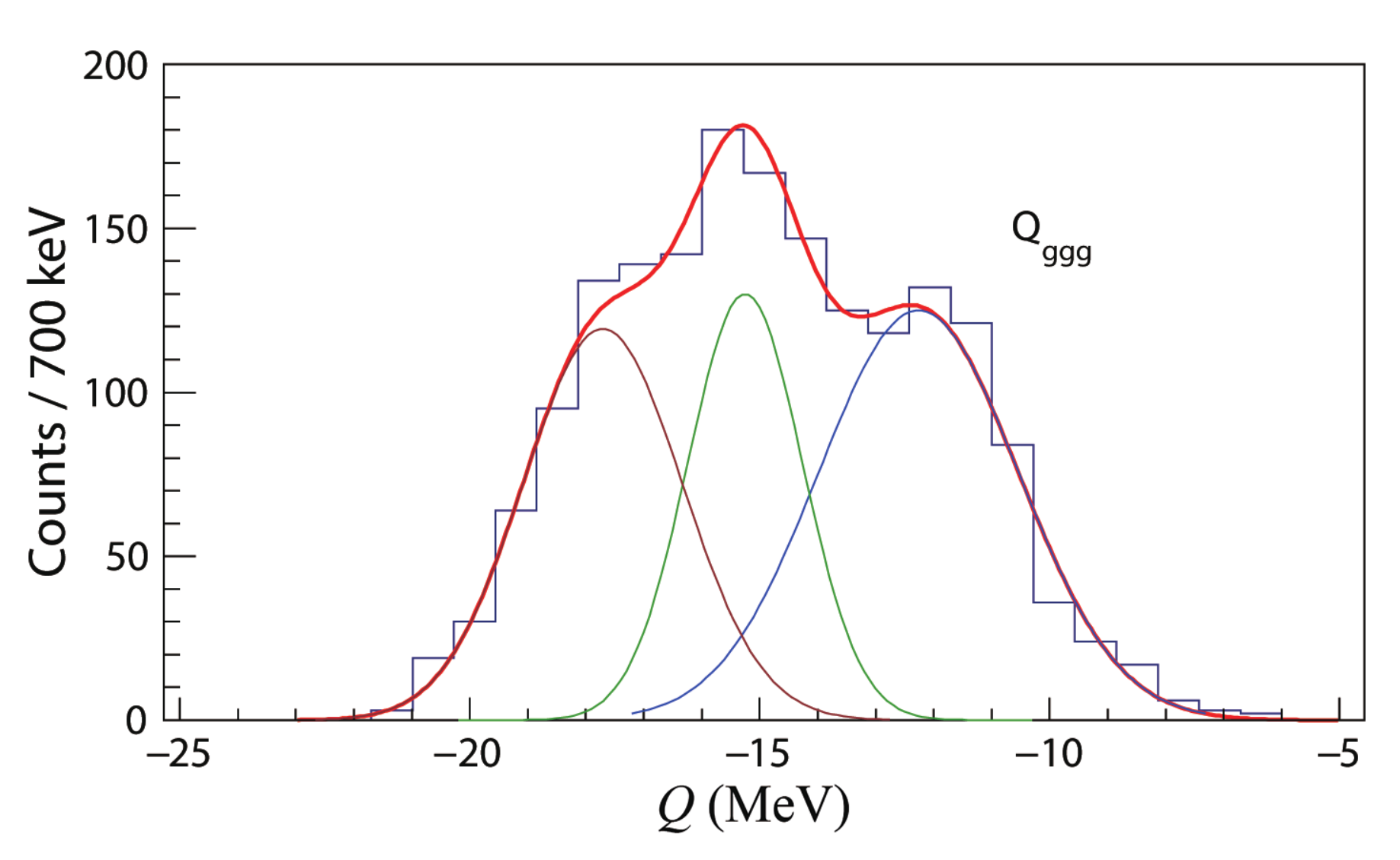}
\figcaption{\label{Q} $Q$-value spectrum for the $^{14}$C $\rightarrow$ $^{10}$Be + $\alpha$ breakup reaction on a proton target, obtained from the present measurement. The $Q_{\rm{ggg}}$ peak is related to the final particles all on their g.s.. The other two peaks correspond to the exit $^{10}\rm{Be}$ in its first excited state (3.36 MeV) and its four adjoining excited states around 6\,MeV.}
\end{center}

The relative energy ($E_{\rm{rel}}$) of the decay fragments $^{10}\rm{Be}+\alpha$ can be calculated from their kinetic energies ($T_{\rm{a}}$ , $T_{\rm{b}}$) and the opening angle $\theta$, according to the invariant mass (IM) method\,\cite{Yang2015}, which takes part in the excitation energy $E_{\rm{x}}$:
\begin{eqnarray}
E_{\rm{x}} &=& E_{\rm{rel}}+E_{\rm{thr}} \nonumber \\
E_{\rm{rel}} &=& \sqrt{M^2}-M_{\rm{a}}-M_{\rm{b}} \nonumber \\
M^2  &=& M^2_{\rm{a}}+M^2_{\rm{b}}+2(M_{\rm{a}}+T_{\rm{a}})(M_{\rm{b}}+T_{\rm{b}}) \nonumber \\
     & & -2\sqrt{(T^2_{\rm{a}}+2T_{\rm{a}}M_{\rm{a}})(T^2_{\rm{b}}+2T_{\rm{b}}M_{\rm{b}})}\cos\theta
\end{eqnarray}
where $E_{\rm{thr}}$ is the separation energy of the $^{14}\rm{C}\rightarrow^{10}\rm{Be}+\alpha$ process, being 12.01 MeV. By gating on the $Q$-value peaks corresponding to the g.s., the first excited state and the states around 6 MeV in the final nucleus $^{10}\rm{Be}$, the $^{14}\rm{C}$ excitation energy spectra can be obtained, as presented in  Fig.~\ref{IM}(a), (b) and (c), respectively. Each spectrum is fitted by a series of Gaussian functions together with a smoothing background function \cite{Curtis1996}. The initialization of the fitting parameters were mostly based on the previously reported results \cite{Soic2003,Li2017}. The $\chi^2$-minimization method was used to determine the center position, width and height of each peak \cite{Li2017}. The extracted resonant energies are listed in Table~\ref{tab1}. The results obtained from previous experiments \cite{Soic2003,Price2007,Li2017}, which applied similar silicon-detectors to measure the decay fragments, are also listed in the table for comparison.
\begin{center}
\includegraphics[width=6cm]{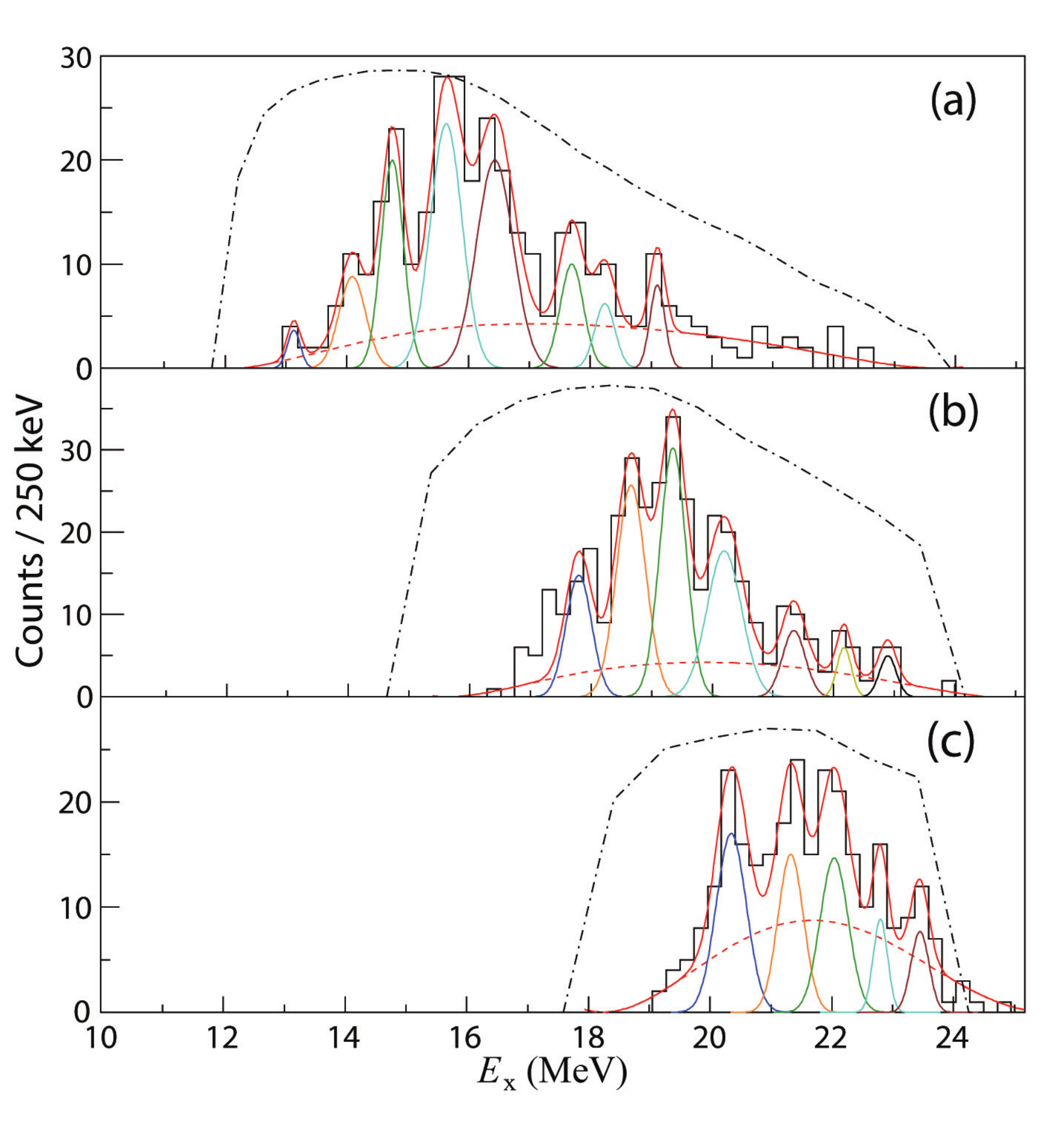}
\figcaption{\label{IM}$^{14}\rm{C}$ excitation-energy spectra reconstructed from the $^{10}\rm{Be} + \alpha$ decay channel and gated on the $^{10}\rm{Be}$ final state according to the $Q$-spectrum in Fig.~\ref{Q}. (a) gated on the $Q_{\rm{ggg}}$ peak at about -12.0 MeV for $^{10}$Be on the g.s.; (b) gated on the peak at about -15.2 MeV for $^{10}$Be on the first excited state at 3.36 MeV; (c) gated on the peak at about -17.9 MeV for $^{10}$Be on the excited states at $\sim$\,6 MeV. Each spectrum is fitted with several Gaussian functions (solid line) and a smoothing background (dotted line). The dot-dashed curve in each spectrum gives the relative detection efficiency.}
\end{center}

Peaks appeared at 14.8, 15.6, 16.4, 17.8, 18.6, 19.1, 20.2, 21.4, 22.2 and 22.8 MeV in current spectrum are all observed in the previous experiments \cite{Soic2003,Price2007,Li2017}, confirming the correctness of the present measurement and the reconstruction analysis. It would be worth noting that the previously observed broad peak at 17.9 MeV \cite{Li2017} is now identified as two peaks at 17.8 MeV and 18.3 MeV, with the latter one being a newly observed state which decays almost exclusively to the g.s. of $^{10}$Be. In addition, the presently observed decay patterns of the states at 18.6, 19.1 and 20.2 MeV exhibit deferent characters compared with previous results. These may be attributed to the different excitation dynamics involved in the corresponding reactions \cite{Oertzen2006}, and be used to selectively constrain the theoretical calculations.

The dot-dashed lines in Fig.~\ref{IM} represent the relative detection-efficiency for three cases of the $^{10}$Be final-state. These were determined from the Monte Carlo simulation using the realistic detection-system setup and the detector's properties. The high detection efficiency at energies close to the decay threshold (12.01 MeV), as emphasized in the present experimental design, is evidenced.

\end{multicols}\begin{center}
\tabcaption{ \label{tab1}  Summary of the resonant states populated in $^{14}\rm{C}$ via the $p(^{14}\rm{C}$,$^{14}\rm{C}^{*}\rightarrow^{10}\rm{Be}+\alpha)$$p$ reaction. For comparison previous results associated with the $\alpha$-decay measurement are also listed in the table\,\cite{Soic2003,Price2007,Li2017}.}
\footnotesize
\begin{tabular*}{170mm}{@{\extracolsep{\fill}}ccccccccc}

\toprule

\multicolumn{3}{c}{This work} & \multicolumn{3}{c}{$^{7}\rm{Li}(^{9}\rm{Be},^{10}\rm{Be}+\alpha)$$d$\,\cite{Soic2003} or $^{14}\rm{C}$ inelastic\,\cite{Price2007}  } & \multicolumn{3}{c}{$^{9}\rm{Be}(^{9}\rm{Be},^{10}\rm{Be}+\alpha)\alpha$\,\cite{Li2017}}\\
$^{10}\rm{Be}_{\rm{gs}}$ & $^{10}\rm{Be}(2^{+})$ & $^{10}\rm{Be}(\sim6\,MeV)$ & $^{10}\rm{Be}_{\rm{gs}}$ & $^{10}\rm{Be}(2^{+})$ & $^{10}\rm{Be}(\sim6\,MeV)$ & $^{10}\rm{Be}_{\rm{gs}}$ & $^{10}\rm{Be}(2^{+})$ & $^{10}\rm{Be}(\sim6\,MeV)$ \\

\hline
13.1(1) &         &         &             &         &         &         &         &         \\
14.1(1) &         &         & 14.3(1)$^*$ &         &         &         &         &         \\
14.8(1) &         &         & 14.7(1)     &         &         &         &         &         \\
15.6(1) &         &         & 15.5(1)     &         &         &         &         &         \\
16.4(1) &         &         & 16.4(1)     &         &         & 16.5(1) &         &         \\
        &         &         & 17.3(1)$^*$ &17.3(1)$^*$&       &         &         &         \\
17.8(1) & 17.9(1) &         &             &         &       & 17.9(1) &         &         \\
18.3(1) &         &         &             &         &         &         &         &         \\
        & 18.6(1) &         & 18.5(1)     & 18.5(1) &         & 18.8(1) &         &         \\
19.1(1) & 19.3(1) &         &             & 19.1(1) &         &         &         &         \\
        &         &         & 19.8(1)$^\triangle$ & 19.8(1) & & 19.8(1) & 19.8(1) &         \\
        & 20.2(1) & 20.3(3) &             &         &20.4(1)$^*$&       & 20.3(1) &         \\
        &         &         & 20.6(1)$^\triangle$ & &20.9(1)$^*$&       &         &         \\
        &         &         &             &         &         & 20.8(1) & 20.8(1) &         \\
        & 21.4(1) & 21.3(3) &             & 21.4(1) &         &         & 21.4(1) & 21.6(3) \\
        & 22.2(1) & 22.0(3) &             &         &         &         & 22.0(1) & 22.0(3) \\
        & 22.8(1) & 22.8(3) &             &         & 22.4(3) &         & 22.5(1) & 22.5(3) \\
        &         & 23.5(3) &             &         &         &         & 23.5(1) & 23.6(3) \\
\bottomrule
\end{tabular*}%
\end{center}
note: in columns 4-6, states with $^*$ or $^\triangle$ were reported only in \cite{Price2007} or in \cite{Soic2003}, respectively, while others were presented in both articles.
\begin{multicols}{2}

\section{States around 14 MeV excitation energy}
As indicated in the introduction, the low-spin states just beyond the $^{14}\rm{C} \rightarrow ^{10}\rm{Be} + \alpha$ separation threshold (12.01 MeV) and with strong cluster-structure are of particular importance for validating the molecular rotational band associated with the $\pi$-bond linear-chain configuration. Previously, the inclusive neutron-transfer reactions, which is in favor of the single-particle-state formation, have strongly populated the states at 12.9 MeV and 14.8 MeV \cite{Oertzen2004,Oertzen2006}. These two states were determined to possess dominantly single-particle structures with spin-parities of $3^-$ and $5^-$, respectively \cite{Oertzen2004,Haigh2008}. In this energy range, another two states at $\sim$\,14.1 MeV and $\sim$\,15.6 MeV behave quite differently. They are populated very weakly in neutron-transfer reactions but much more significantly in multi-hole multi-particle  transition processes, the latter being known as cluster-creation tools \cite{Oertzen2004, Oertzen2006}. Their resonance-widths are also relatively broad with respect to those of typical single-particle states \cite{Oertzen2004}. Indeed, based on the $^{14}\rm{C} \rightarrow ^{10}\rm{Be} + \alpha$ decay analysis, the state at 15.6 MeV was identified as a typical cluster resonant state with spin-parity of $3^-$ \cite{Price2007,Haigh2008}. It was difficult to carry on the same cluster-decay analysis for the $\sim$\,14.1 MeV state since the detection of the near-threshold cluster-emission is much more difficult \cite{Price2007}.

In the present work, we focused on the improvement of the detection of the $\alpha$-cluster decay from states in $^{14}$C around 14 MeV excitation energy. The special difficulties here are threefold. Firstly, the population of the low-spin states may have smaller cross section according to the approximate ($2J + 1$) proportional rule \cite{Oertzen2004}. Secondly, the charged-cluster decay would be largely suppressed for states close to the decay-threshold, due to the relatively large Coulomb barrier. And thirdly, the small decay-energy combined with low-spin means that both decay-fragments may emit at very forward angles \cite{Yang2014}. We therefore have applied the inverse kinematics in the reaction, together with the 0-degree detection system, in order to have an almost 100\,$\%$ detection probability for the cluster-decay events \cite{Yang2014, Yang2015}. In Fig.~\ref{IMwithFreer}, the previously reported resonances, reconstructed from the $^{14}$C $\rightarrow$ $^{10}$Be\,(g.s.) + $\alpha$ process, are compared with the present observation, together with the detection efficiency curves. It is clear that the present detection system has an obviously superior coverage of the near-threshold energy range. In both Fig.~\ref{IMwithFreer}(a) and (b), we see clear identifications of the resonances at 14.8 and 15.6 MeV, which were previously determined to have spin-parities of $5^-$ and $3^-$, respectively. In addition, a hint of resonance at 14.3(1) MeV initially seen in \cite{Price2007} (Fig.~\ref{IMwithFreer}(a)) is now clearly observed as a peak at 14.1(1) MeV (Fig.~\ref{IMwithFreer}(b)), thanks to the near-threshold high detection efficiency of the present measurement. The number of counts for the 14.1 MeV resonance is 19, corresponding to a significance of observation certainly higher than $3\sigma$, after taking into account the background fluctuation. The relatively small $\alpha$-decay yield of the 14.1 MeV state in $^{14}$C, compared with that for the 14.8 or 15.6 MeV state, is reasonable considering the effect of the Coulomb barrier which is much stronger for states closer to the decay threshold.
\begin{center}
\includegraphics[width=6cm]{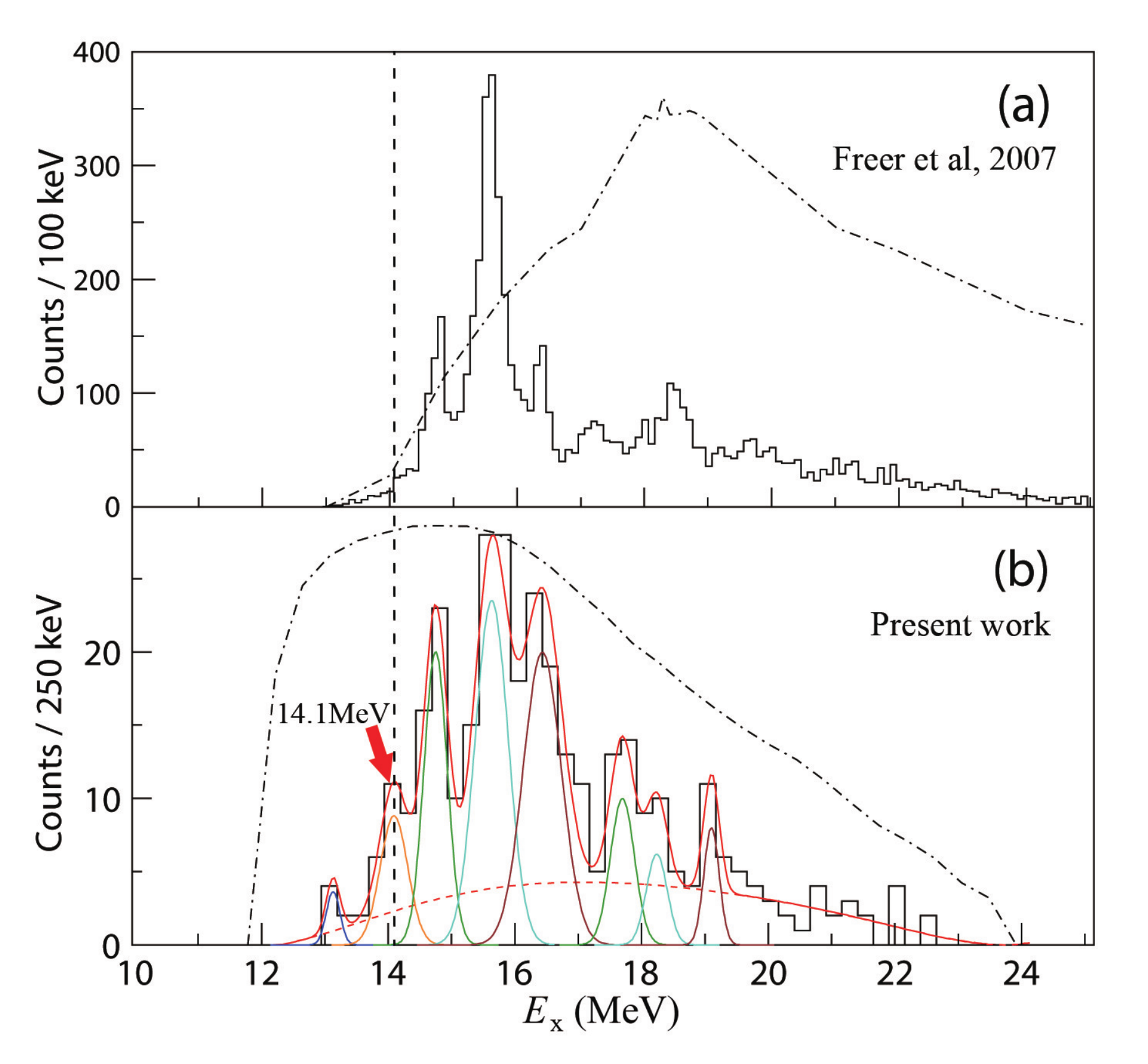}
\figcaption{\label{IMwithFreer} Resonant states reconstructed from the $^{14}$C $\rightarrow$ $^{10}$Be\,(g.s.) + $\alpha$ breakup events, measured in the previous (a) and in the present (b) experiments. The dot-dashed curves represent the corresponding relative detection efficiencies. The vertical black-dashed line is used to indicate the position of the 14.1 MeV state.}
\end{center}

Due to the limited number of counts, it would be difficult to make direct spin-determination analysis, e.g. by using the angular correlation \cite{Freer1996,Price2007,Yang2015} or the differential cross section methods \cite{Price2006}. However, we may still use the observed relative yields to estimate the possible spin assignment. Here we use the state at 15.6 MeV as a reference, which behaves like a pure clustering resonance with a well determined spin-parity of $3^-$ \cite{Price2007}. As indicated above, the population and cluster-decay properties of the 14.1 MeV state is similar to those of the 15.6 MeV state \cite{Oertzen2004}, and therefore the quantitative yield comparison between them should be reasonable. The $\alpha$-decay yield from a resonant state can be expressed as
\begin{eqnarray}\label{yield}
%k &=& \sqrt{2\mu E} \nonumber \\
N_{\rm{exp}}(E) &=& IT \sigma \varepsilon R_\alpha(E),
\end{eqnarray}
where $I$ is the number of incident beam particles, $T$ the target thickness, $\sigma$ the reaction cross section, $\varepsilon$ the detection efficiency and $R_\alpha$ the $\alpha$-decay branching ratio (BR) of the resonance. According to the $2J+1$ rule, for $J$ being the spin of the resonant state, we have approximately a reaction cross section of $\sigma = \sigma_0 (2J+1)$\,\cite{Oertzen2004}. The ratio $R$ can be deduced from $\Gamma_\alpha / \Gamma_t$, with $\Gamma_\alpha$ and $\Gamma_t$ the partial $\alpha$-decay width and the total decay width, respectively, of the resonance. Based on the single-channel R-matrix approach \cite{Lane1958,Yang2014b}, the partial decay width $\Gamma_{\alpha}(E)$ is related to the dimensionless reduced width (the cluster spectroscopic factor) $\theta_{\alpha}^2$ :
\begin{eqnarray}\label{SP}
 \Gamma_{\alpha}(E) &=& 2 \theta_{\alpha}^2 \gamma_{W}^2 P_{l}(E)
\end{eqnarray}
with the Coulomb barrier penetrating factor $P_{l}(E)$ and the cluster-structure Wigner Limit $\gamma_{W}^2$ defined as ~\cite{Lane1958,Yang2014b}:
 \begin{eqnarray}\label{P}
  P_{l}(E)  &=& \frac{ka}{(F_l(ka))^2+(G_l(ka))^2}  \nonumber  \\
  \gamma_{W}^2 &=& \frac{3\hbar^2}{2\mu a^2}.
\end{eqnarray}
Here $F_l(ka)$ and $G_l(ka)$ are the standard regular and irregular Coulomb wave functions, respectively. The parameters $E$, $l$, $\mu$ and $k=\sqrt{2\mu E}$ are the relative energy, the relative orbital angular momentum, the reduced mass and the wave number, respectively, of the decay partners. For the system of $^{10}\rm{Be}\,(g.s.) + \alpha$, $J = l$ is valid since both final fragments have spin-zero. The $P_{J}(E)$ and $\gamma_{W}^2$ can then be calculated exactly by using the channel radius $a = r_0(10^{\frac{1}{3}}+4^{\frac{1}{3}})$ with $r_0\approx1.4$ fm. Since $I$, $T$, $\sigma_0$ and $\varepsilon$ may be treated as constants for 14.1 and 15.6 MeV states in the same reaction, the yield ratio of these two states can be expressed in the form:
\begin{eqnarray}\label{ratio}
\frac{N_{\rm{exp}}(14.1)}{N_{\rm{exp}}(15.6)} =\frac{\theta_{\alpha}^2(14.1) \Gamma_t(15.6)}{\theta_{\alpha}^2(15.6) \Gamma_t(14.1)}\cdot \frac{(2J_x+1)P_{J_x}(14.1)}{7 P_{3}(15.6)}.
\end{eqnarray}
Here we have taken $J=3$ for the known 15.6 MeV state and $J_x$ as variable for the 14.1 MeV state. The present experimental yield ratio (Fig.~\ref{IMwithFreer}) is $N_{exp}(14.1)/N_{exp}(15.6)$ = 19/61 = 0.31\,($\pm\,0.10$). The calculated value for $P_{3}(15.6)$ is 0.81, and for $P_{Jx}(14.1)$ are 1.14, 0.84, 0.40, 0.10 and 0.015 for $J_x$ equal to 0, 1, 2, 3 and 4, respectively. Based on the existing knowledge (see above discussion) of the 14.1 MeV and 15.6 MeV states, their $\theta_{\alpha}^2$ (or $\Gamma_t$) values are quite similar to each other \cite{Oertzen2004}. In this case, the right side of the Eq.~\ref{ratio} generates values of 0.20, 0.45, 0.36, 0.125 and 0.024 for $J_x$ equal to 0, 1, 2, 3 and 4, respectively. It can be seen that only the first three values are close to the experimental one ($0.31\,\pm\,0.10$), even for a variation of the $\theta_{\alpha}^2$/ $\Gamma_t$ value by a factor of $50\%$. In other words, current relative yields analysis, by taking into account the Coulomb barrier penetrability, tends to constrain the spin of the 14.1 MeV resonance at low values of $0\sim2$, being consistent with the expectation of a $0^+$ band-head at around 14 MeV for the $\pi$-bond linear-chain configuration in $^{14}$C.

We note that the inclusive missing-mass (MM) measurement using the telescope T$_1$ and T$_2$ were not used in the present analysis, due to the energy resolution significantly larger than the required resonance-separation. In principle, the beam-energy spread affects much less the IM reconstruction, as described above, but largely influences the direct MM measurement~\cite{Yang2014}. As a matter of fact, it seems difficult to identify the 14.1 MeV state in MM spectrum due to the relatively small population probability of the exotic clustering state, if normal transfer reaction or inelastic-scattering tools were used ~\cite{Oertzen2004}. This difficulty would prohibit the extraction of the cluster-decay BR from the IM + MM measurements ~\cite{Yang2014b}. However, it was evidenced that the multi-hole multi-particle transition processes are in favor of the population of the 14.1 MeV state ~\cite{Oertzen2004}. This should be considered in the future investigation by applying both the recoil particle (MM) and the cluster-decay (IM) measurements.

\section{Summary}
An inelastic excitation and decay  experiment, $p(^{14}\rm{C}$,$^{14}\rm{C}^{*}\rightarrow^{10}\rm{Be}+\alpha$)$p$, was performed at a beam energy of 25.3 MeV/u. The final states of $^{10}\rm{Be}$ are distinguishable based on the $Q$-value analysis. The excitation energies of the mother nucleus $^{14}$C are reconstructed with respective to the $^{10}\rm{Be}$ on its g.s., first excited state and multiplet around 6 MeV excitation. Most of the observed resonant states in $^{14}$C are in good agreement with previously reported results, confirming the correctness of the present measurement and reconstruction. A previously observed broad state at about 17.9 MeV is now identified as two states at 17.8 MeV and 18.3 MeV, with the latter being a new observation. The decay-patterns of a few states have been supplemented. Most importantly, a state at 14.1 MeV is now clearly identified from the cluster-decay channel, thanks to the specially designed 0-degree detection system. The relative yield of this state, compared with the neighbouring known state, allows to constrain its spin within $0 \sim 2$. This observation and spin-constrain is important since it provides most likely the band-head of the molecular rotation band with the $\pi$-bond linear-chain configuration, which was predicted by the AMD model-calculations and by the experimental-data projections. It would be very interesting to further study this state with direct determination of its spin and cluster-decay branching ratio. Reactions with multi-hole multi-particle excitation would be in favor of observing this state in both inclusive spectrum and in cluster-decay channel. Theoretical investigation of this state is also badly needed.

\acknowledgments{We appreciate the staff of HIRFL-RIBLL in Lanzhou for their excellent work in providing the beams and all kinds of supports. This work is supported by the National Key R$\&$D Program of China (the High Precision Nuclear Physics Experiments) and the National Natural Science Foundation of China (Nos. 11535004, 11775004, 11405005, 11375017).}

\end{multicols}

\vspace{-1mm}
\centerline{\rule{80mm}{0.1pt}}
\vspace{2mm}

\begin{multicols}{2}

\end{multicols}

\clearpage

%\end{CJK*}
\end{document}